# Quantum sensing of motion in colloids via time-dependent Purcell effect

Alexey S. Kadochkin[1,2,*], Ivan I. Shishkin[3,4], Alexander S. Shalin[2],

and Pavel Ginzburg[3,4]

[1]Ulyanovsk State University, Ulyanovsk, 432017, Russia
[2]"Nanooptomechanics" Laboratory, ITMO University, St. Petersburg, 197101, Russia
[3]School of Electrical Engineering, Tel Aviv University, Tel Aviv, 6997801, Israel
[4]Light-Matter Interaction Centre, Tel Aviv University, Tel Aviv, 6997801, Israel

Keywords: Purcell effect, Local field enhancement, Spontaneous emission, Mesoscopic Models, Fluid flow, Nanoparticles, Phosphorescence

**Abstract:**

Light-matter interaction dynamics is governed by the strength of local coupling constants, tailored by surrounding electromagnetic structures. Characteristic decay times in dipole-allowed fluorescent transitions are much faster than mechanical conformational changes within an environment and, as the result, the latter can be assumed static during the emission process. However, slow-decaying compounds can break this commonly accepted approximation and introduce new interaction regimes. Here, slow decaying phosphorescent compounds are proposed to perform quantum sensing of nearby structure's motion via observation of collective velocity-dependent lifetime distributions. In particular, characteristic decay of an excited dye molecule, being comparable with its passage time next to a resonant particle, is modified via time-dependent Purcell enhancement, which leaves distinct signatures on properties of emitted light. Velocity mapping of uniformly moving particles within a fluid solution of phosphorescent dyes was demonstrated via analysis of modified lifetime distributions. The proposed interaction regime enables performing studies of a wide range of phenomena, where time-dependent light-matter interaction constants can be utilized for extraction of additional information about a process.

* Corresponding author: askadochkin@sv.ulsu.ru

## Introduction

Quantum laws of nature open a room of opportunities to enable numerous practical applications, where superior sensing performances are required [1],[2]. Exploration of quantum properties of light is just one example among many others. While a majority of the proposed schemes relies on far-field illuminations (e.g. ghost imaging [3]), investigation of near field interactions enables extraction of the information about nanoscale phenomena, as it will be shown hereafter. Properties of a spontaneous emission tailored by nearby electromagnetic environment will be employed to monitor a motion of the latter.

Acceleration of spontaneous emission rates in structured environment respectively to a free space is called Purcell effect [4]. While traditional pathways towards accelerating radiative decays and even reaching strong coupling regimes of interaction utilize high quality factor cavities, recent advances in nanofabrication offer complementary solutions of using small subwavelength resonators, e.g. [5],[6],[7]. This approach is based on open cavities and suggests manipulation of light-matter interaction processes via local field control or, in other words, via small modal volumes [8],[9],[10],[11],[12]. For example, metallic nanoantennas supporting localized plasmonic resonances were shown to provide flexible solution for achieving moderate Purcell enhancement and directionality in emission, e.g. [8],[10],[13]. In contrary to traditional cavities case, the near-field coupling approach requires emitter placement in a close vicinity to antenna boundaries. Purcell factor enhancement is also used for increasing the brightness of fluorescence probes in imaging applications [14],[15]. It is worth noting that characteristic lifetimes of dipole-allowed fluorescent processes lie within nanosecond scale, making any conformational changes within a surrounding environment to be virtually time-independent. While different microscopy techniques to monitor dynamical changes at real time, such as fluorescent lifetime imaging (FLIM) [16] or gated photoluminescence microscopy [17] do exist, they assume that the environment is at the instantaneous rest. Here a new regime of time-dependent light-matter interaction in application to the motion detection in fluids will be investigated.

Information on a particle motion within a solution is indispensable in many disciplines including chemistry, biology and medicine. "Lab on a chip" applications require reliable and accurate monitoring of fluid flow with high spatial resolution [18]. In many cases, velocities of a fluid and a particle within it can differ from each other and the information on the relative motion is essential. Many methods for monitoring this property have been developed, e.g. [19], [20], [21], [22], [23], [24],[25]. In particular, external partially transparent masks and fluorescent flow tracers were employed in [19], [20]. The method reported in [21] utilizes the cross-correlation analysis in time-dependent optical speckle field, where the velocity is extracted via the perturbation introduced by the particle. Image velocimetry technique is introduced in [22] and it relies on cross-correlation post-processing of fluorescent signals. Velocimetry based on the direct imaging of fluorescent particles in a flow is also employed in [23], where several sequential camera snapshots are analysed. Doppler spectroscopy can be also employed for mapping relatively high velocities and it requires quite an expensive apparatus [24]. It is worth noting, however, that all of the beforehand mentioned methods do not allow velocity mapping deeply beyond the diffraction limit and, in many cases, require complex measurement setups. Here a qualitatively different method relying on quantum nature of spontaneous emission of light is proposed and analysed.

Fluid flow with an assembly of dissolved slow decaying luminescent tags around a resonant nanoparticle will be considered. The decay rate of the tags is chosen to be comparable with their time of flight next to the particle, which is assumed to be immobilized. As the result, the local density of states, which manifests itself via instantaneous emission rate, will change during the emission time. Consequently, the law of luminescence decay will inherit the information regarding the relative velocity of the fluid. Here, the phosphorescent and rare earth-based luminescent tags, having their lifetimes in micro-and millisecond ranges, will be considered analytically. The lifetime distributions acquired from the analysis of the collective decay will be mapped on the velocity field. The proposed method offers distinguished advantages towards nanoscale minimally invasive mapping of flows over the large span of probable velocities, as it will be demonstrated.

The typical scenario, that will be studied hereafter, is depicted in Fig. 1. The manuscript is organized as follows: theoretical framework of spontaneous emission in a time-varying environment will be developed at first and then applied on an example of laminar fluid flow around a nanoantenna. Finally, a realistic scenario will be analyzed with the developed formalism and the possibility of accurate mapping of velocities relevant to micro and nano fluidic applications will be demonstrated.

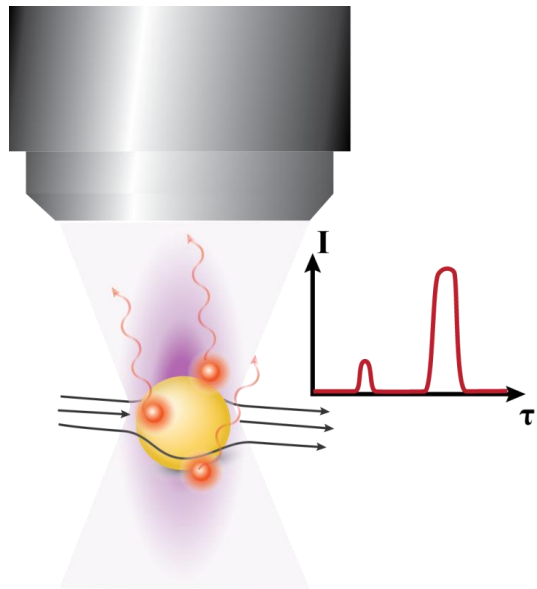

**Fig. 1**. Schematics of the setup layout – a flow of luminescent tags around a nanoantenna is detected via modified spontaneous emission rates analyzed with the lifetime distribution postprocessing.

**Lifetime distribution analysis of spontaneous emission rates in an environment with time-dependent Purcell enhancement**

Spontaneous emission dynamics can be described with the help of semi-classical rate equations, in the case, when quantum coherence effects are neglected [26]. Here, the time-dependent coupling coefficient proportional to Local Density of States (LDOS) will be introduced within this framework. It is worth briefly noting the existence of several quantum electrodynamical models that introduce mechanical degrees of freedom explicitly into the field quantization, e.g. [27],[28],[29], but they cannot address complex geometries explicitly. Furthermore, the interaction regime considered here should not be confused with the case of cavity quantum optomechanics [30], where the back action of the emission on mechanical motion is the key parameter. Strong decoherence effects in a majority of room temperature scenarios, especially in fluidics, make the consideration of back action effects redundant on one hand, but allows analyzing complex practical scenarios on another, as the one considered hereafter.

LDOS has a well-defined physical interpretation in the case of the weak coupling. The conditions for such interaction regime between phosphorescent molecules and the metallic particle are fulfilled if separation distances larger than 10 nm. As the result, the main contribution to the collective luminescent time-dependent signal will be obtained from the volume, where quantum revivals can be neglected. Furthermore, consideration of separation distances smaller than 10 nm requires extra-care, since quenching mechanisms start playing a role, e.g.[31],[32],[33].

The following analysis is performed for the flow of an incompressible infinitesimal volume of solution with phosphorescent dyes propagating along a streamline. Subsequently, volumetric summation over the total amount of molecules will be performed in order to extract time-dependent macroscopic luminescent signal. It is worth noting that the proposed analysis could be applied to other time dependent processes and not only to fluid dynamics. The decay of the molecules situated within an infinitesimal volume $dV_0$ is described by the following rate equation:

$$\frac{dN(\mathbf{r}(r_0,\theta_0,\varphi_0,t))}{dt} = -P(\mathbf{r}(r_0,\theta_0,\varphi_0,t)) \cdot \gamma_0 \cdot N(\mathbf{r}(r_0,\theta_0,\varphi_0,t)), \qquad (1)$$

where $\gamma_0$ is the decay rate of molecules in a free space, $N(\mathbf{r}(r_0,\theta_0,\varphi_0,t))$ is the time-dependent density of excited molecules within the infinitesimal volume $dV_0$, $\mathbf{r}(r_0,\theta_0,\varphi_0,t)$ is the trajectory of the infinitesimal volume as the function of time with $(r_0,\theta_0,\varphi_0)$ being its initial location at the time of a pump pulse arrival, and $P(\mathbf{r}(r_0,\theta_0,\varphi_0,t))$ is the position-dependent Purcell factor, averaged over spatial orientations of the radiating dipole. In the case of a spherical particle, the orientation-averaged Purcell factor depends only on the distance between the test volume and the center of the particle. Eq. 1 has a solution in the form of:

$$N(\mathbf{r}(r_0,\theta_0,\varphi_0,t)) = N(\mathbf{r}(r_0,\theta_0,\varphi_0,0)) \cdot \exp(-\int_0^t P(r(r_0,\theta_0,\varphi_0,t'))) \cdot \gamma_0 \cdot dt'), \quad (2)$$

where $N(\mathbf{r}(r_0,\theta_0,\varphi_0,0))$ is the initial density of excited molecules in space (note, that this quantity is non-uniform due to the antenna effect of the particle acting on the pump field). Thus, the collective time-dependent fluorescent signal can be obtained knowing the decay rate of the molecules. However, an additional correction factor should be introduced in order to take into account the nonradiative quenching and light propagation effects. The following relation gives the infinitesimal contribution to the luminescent signal intensity:

$$dI(\mathbf{r}(r_0,\theta_0,\varphi_0,t)) = P_{rad}(r(r_0,\theta_0,\varphi_0,t)) \cdot \gamma_0 \cdot C_{eff}(r(r_0,\theta_0,\varphi_0,t)) \cdot N(\mathbf{r}(r_0,\theta_0,\varphi_0,t)) dV_0, \quad (3)$$

where $P_{rad}(r(r_0,\theta_0,\varphi_0,t))$ and $C_{eff}(r(r_0,\theta_0,\varphi_0,t))$ are position-dependent radiative enhancement and collection efficiency respectively – both are averaged over the orientation of a dipole moment within an environment. While the overall population inversion law is governed by the total radiative decay (Eq. 1), those two factors account for photon loss, originating from both nonradiative quenching and imperfect collection efficiency. The later factor strongly depends on the implementation of an experimental setup (numerical aperture of the collection optics + the shape of the nanoantenna) and will be taken to be unity for subsequent theoretical analysis. Nonradiative quenching is calculated by applying semi-classical approaches, relying on the comparison between Poynting vector flux calculations in the vicinity of the emitter and in the far field [34],[35].

In order to obtain the contribution of all infinitesimal volumes $dV_0$, initially located at $(r_0,\theta_0,\varphi_0)$ at $t=0$, intensity decay law of Eq. 3 should be integrated over the whole space $V_0$:

$$I(t) = \int_{V_0} dI(\mathbf{r}(r_0,\theta_0,\varphi_0,t)). \quad (4)$$

The developed approach towards the collective quantum sensing of the flow in the fluid environment will be tested on a system with known passage trajectory $\mathbf{r}(r_0,\theta_0,\varphi_0,t)$ of each of the infinitesimal volume $dV_0$ in the next Section. The lifetime distribution analysis will be performed by applying inverse Laplace transform over the time-dependent intensity.

**Laminar flow around a plasmonic spherical nanoparticle**

The laminar flow around a nanoscale spherical particle will be considered. In the case of low Reynolds numbers [18],[36] (*Re*<<1 , Stokes flow) the flow has a closed form analytical description. The velocity field in the spherical coordinate system is given by the following relation [36]:

$$\mathbf{u} = v_0\left(1-\frac{a^3}{r^3}\right)\cos\theta\,\mathbf{e}_r - v_0\left(1+\frac{a^3}{2r^3}\right)\sin\theta\,\mathbf{e}_\theta, \qquad (5)$$

where $a$ is the radius of the particle situated at the origin, $r$ and $\theta$ are coordinates in the spherical coordinate system, $\mathbf{e}_r$ and $\mathbf{e}_\theta$ are unit vectors along directions of $r$ and $\theta$ correspondingly, and $v_0$ is the fluid velocity far away from the particle (z-axis is directed towards the fluid flow). The coordinates of the infinitesimal volume of the fluid in Lagrange variables is given by the following time derivatives [36]:

$$\frac{dr}{dt} = v_0\left(1-\frac{a^3}{r^3}\right)\cos\theta, \qquad (6)$$

$$r\frac{d\theta}{dt} = -v_0\left(1+\frac{a^3}{2r^3}\right)\sin\theta.$$

The solution of the Eq. 6 enables extracting the path $\mathbf{r}(r_0,\theta_0,\varphi_0,t)$, which is required for the evaluation of the integral time-dependent intensity (Eq. 4).

In the following model system, the immobilized gold spherical particle of 10 nm diameter was chosen in order to demonstrate the performance of the developed method. A static system of an emitter next to a spherical resonator was extensively studied, e.g. [37], and radiation properties of such a system are well understood. It is worth noting that low-complexity geometry was chosen on purpose to simplify the mathematical analysis, though the model could be extended to the more complex cases. Green's functions analysis (either analytical or numerical) enables extracting all the coefficients required for evaluating the expression in Eq. 4.

The time-dependent intensity is studied under the pulsed excitation. Thus, the number of activated dye molecules at the first instance of time (after deactivation of the excitation beam) is proportional to the pump field intensity under the condition of the weak non-saturating excitation. The cases of two orthogonal polarizations of the pump, namely along and perpendicular to the fluid flow, can be identified owing to the evolution of the population inversion in time. The intensity decay rates fall within one of the three major regimes, which can be distinguished by comparing the total radiative lifetime in the vicinity of the particle with the time of flight of a molecule next to the structure. Hereafter, the following dimensionless number corresponding to the ratio between the two times will be used (normalized time of flight):

$$P_{flight} = \frac{v_0}{\pi R \gamma_0 \langle P \rangle} \qquad (7)$$

where $\langle P \rangle$ is the averaged Purcell factor at the dipolar modal volume of the antenna, and $\pi R$ is the half of the particle's circumference (in general, a geometrical parameter of an antenna). In case $P_{flight} \ll 1$ the relative flow of the dyes around the particle is very slow and their excitation will decay much faster than any conformational change in space would occur. $P_{flight} \gg 1$ represents the case of fast fluid flows when the molecules almost do not interact with the particle. The most illustrative, but not necessarily most appropriate for the velocimetry regime for the detection of a motion is the case of $P_{flight} \sim 1$ - here the time-changing environment has the most distinguishable signature on the lifetime.

Figs. 2 and 3 summarises the obtained simulation results for the beforehand mentioned flow regimes for two orthogonal polarizations of the excitation. Instantaneous emission intensity (Eq. 3) appears in the colour maps. The left columns on those figures correspond to the regime of $P_{flight} \ll 1$ - here the particle causes the fast relaxation, and the cloud of the depleted population inversion goes down the streamlines. The right columns demonstrate the detachment of the population inversion from the nanoparticle. The regime of $P_{flight} \sim 1$ (the column at the middle) clearly demonstrates the combined behaviour. The distinguished difference between the two orthogonal polarizations of the excitation can be seen in the case of long and intermediate normalized passage times. Here, in the case of the polarization parallel to the streamlines (gray lines at Figs. 2 and 3) the left hotspot (upstream) of the population inversion has to travel a longer path around the particle and, as the result, has a longer interaction time with the nanoantenna. In case the relaxation is fast, the polarization sensitivity of the flow is less pronounced (left columns on Figs. 2 and 3). In fact, the regime of $P_{flight} \ll 1$ is the most suitable one for velocimetry – in this case the volume fraction of the depleted population (the purple cloud on the left column) grows proportionally to the velocity and contribute to the Purcell-modified relaxation components, which can be clearly distinguished by applying the lifetime distribution analysis, as it will be shown hereafter.

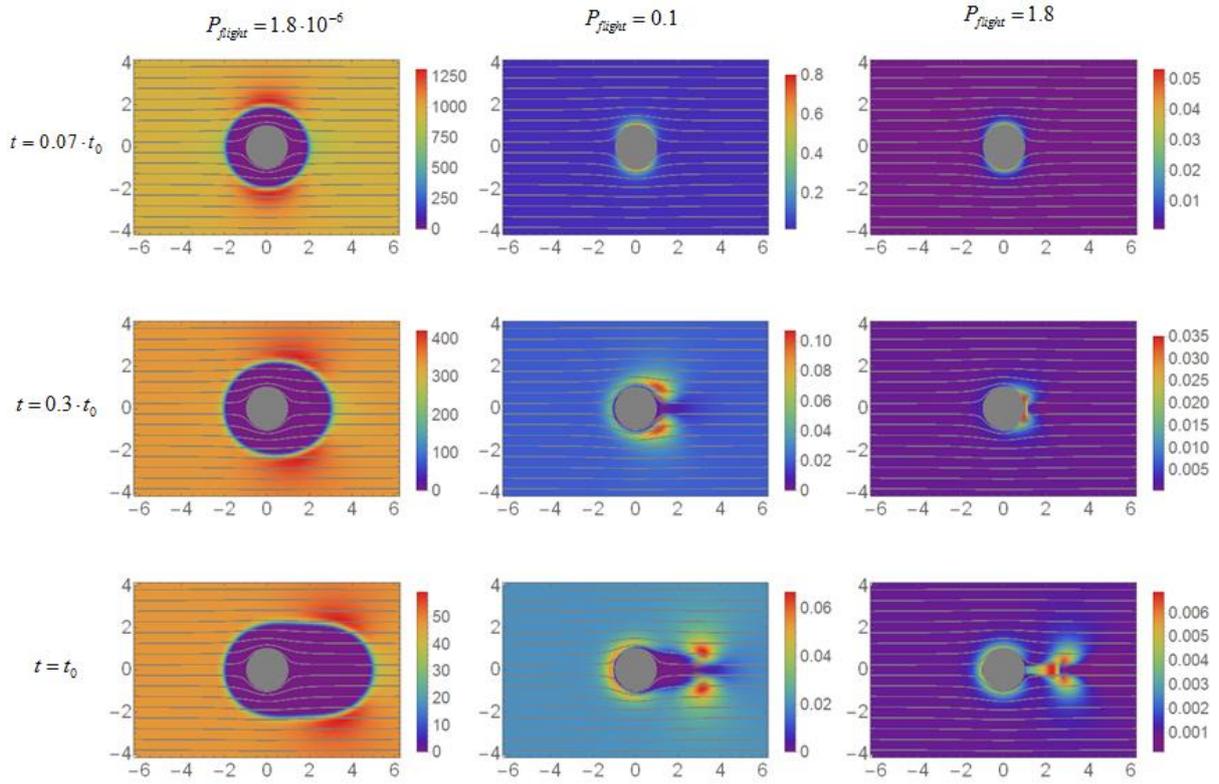

Fig. 2. Snapshots of the instantaneous intensity (Eq. 3) color maps at different instances of time (indicated at the left column). *The pump is polarized perpendicular to the flow.* $t_0 = \pi a/v_0$. x- and y-coordinates are distances normalized to the particle's radius.

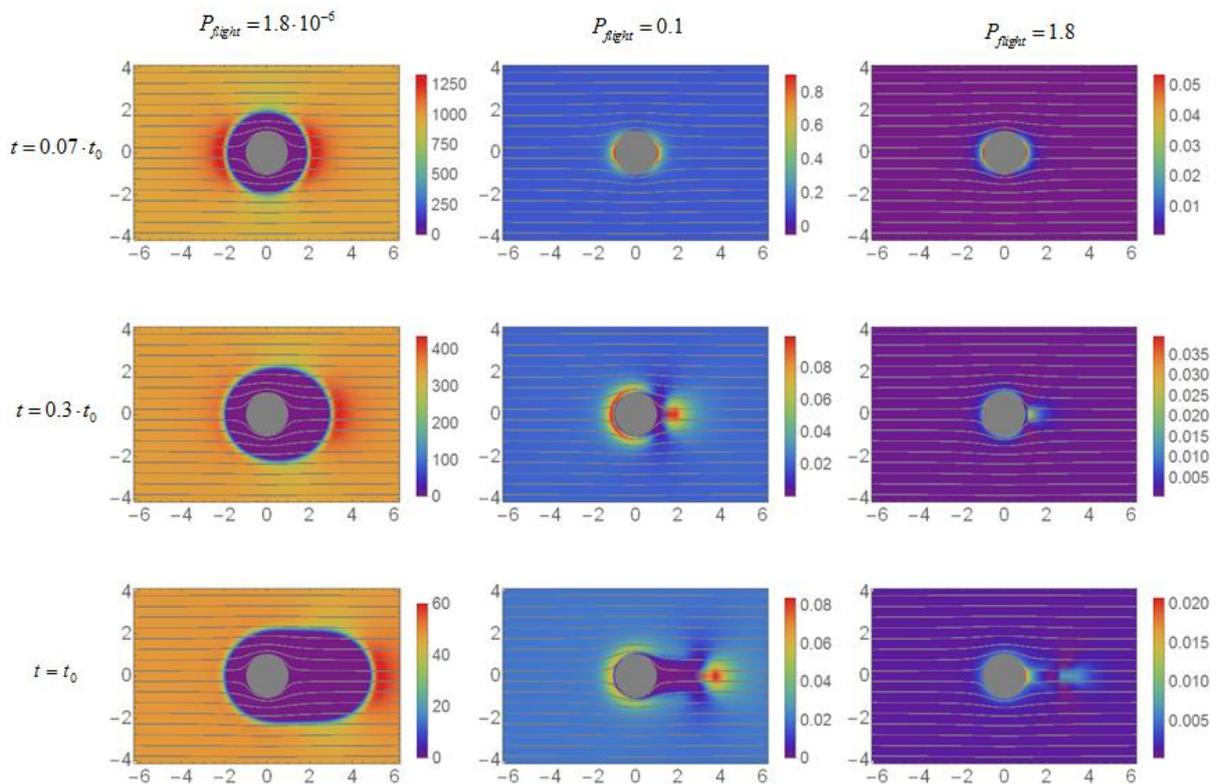

Fig. 3. Snapshots of the instantaneous intensity (Eq. 3) color maps at different instances of time (indicated at the left column). *The pump is polarized parallel to the flow.* $t_0 = \pi a/v_0$. x- and y- coordinates are distances normalized to the particle's radius.

**Lifetime distribution analysis**

While deeply subwavelength motion of the excitation cloud cannot be imaged directly owing to the diffraction limit, lifetime analysis of the collective signal can provide valuable information. The time trace of the intensity is predominated by the unmodified lifetime of the luminophores, situated apart from the plasmonic particle (far region, discussed before). Owing to the collection from the diffraction limited spot, the volumetric ratio between near and far zones (either affected by the particle or not) is ~1:10. It is worth noting in brief that the additional antenna effect of directive emission can improve this ratio in the favor of lifetime-modified photons (it will come via $C_{eff}$ in Eq. 3). This effect has a minor impact in the case of the spherical antenna and will be ignored here. In order to obtain information about the system behaviour, lifetime distribution analysis will be pursued [38]. The goal is to apply the inverse Laplace transform on the time-dependent intensity function $I(t)$ and identify contributions $g(s)$ of different exponential decays $e^{-st}$ to the overall signal:

$$I(t) = \int_0^\infty g(s)e^{-st}ds. \tag{8}$$

Here $g(s)$ is the relative weight of the exponential contributions and $s = 1/\tau$ with $\tau$ being the relaxation time. In order to extract the information on the lifetime distributions, the integral equation (Eq. 8) should be solved. This procedure was done with the numerical routine, similar to those, reported at [38], [39].

**Detection of a flow with phosphorescent dyes – numerical example**

For the demonstration of the effectiveness of the developed analysis method the evaluation of the lifetime change will be carried for a model system. Relatively large span of phosphorescent (e.g. [40]) and rare earth (e.g. [41]) luminescent molecules with characteristic decay times on the scale of micro to milliseconds can be employed for the velocity mapping. Here Eu$^3$, demonstrating 1 msec lifetime, was taken as a test dye [41]. The pump wavelength, corresponding to the peak absorption, is 346 nm, while the dye emission peak is at 618 nm. The illumination spot is half a wavelength in diameter. Fig. 4 summarizes the results for both polarizations of the excitation. As it can be seen that apart from the main relaxation time contribution of the unmodified lifetimes (around 1 msec), there exists a secondary velocity-dependent peak in the region between $10^{-3} - 10^{-2}$ msec. It is associated with the Purcell enhancement by the particle, which holds the key information on the flow motion. Here the lifetime distribution peak position accurately follows the change in the velocity and enables mapping of the velocity via the lifetime distribution analysis. The position of the relevant peak as the function of the fluid velocity appears in the figure insets. The increase in the fluid velocity causes shortening of the central lifetime of those secondary peaks. The polarization sensitivity of the effect also might enable detection of the flow direction. The detectability area of probable velocities is quite broad and covers 80–1200 µm/sec range for a dye with 1 ms decay time, which is very relevant to many microfluidic applications [19], [22], [42].

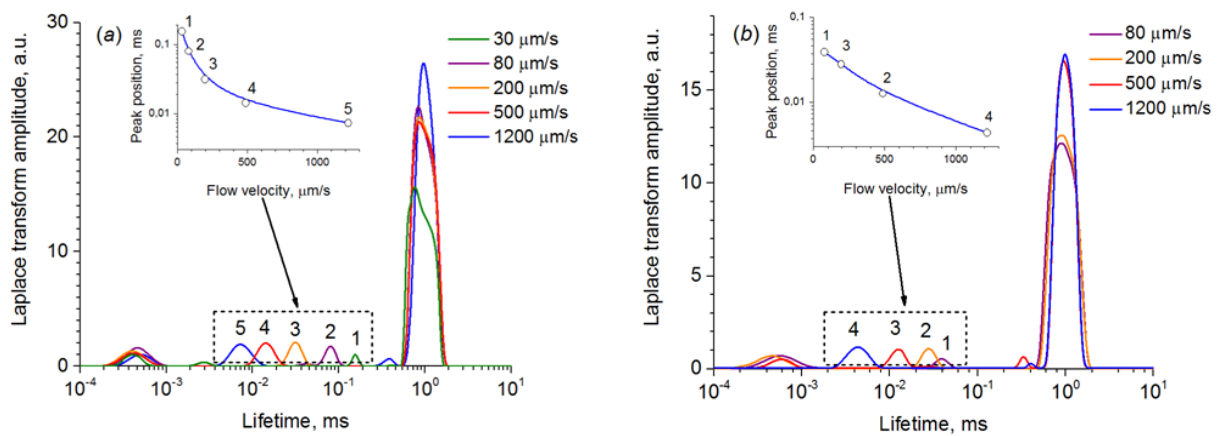

Fig 4. Lifetime distribution analysis of the collective signal. Lifetimes of $Eu^3$ complex dissolved in water excited around 10nm gold sphere. Optical pump is polarized (a) parallel to the flow, (b) perpendicular to the flow. Colour lines correspond to different velocities of the flow (in captions). Insets – position of the secondary peak (in dashed box), as the function of the fluid velocity.

**Conclusions**

Dynamic regime of light-matter interaction with time-varying environment was investigated and applied to the problem of a fluid flow detection beyond the diffraction limit of optical apparatus. Collective decay of the dye molecule ensemble flowing in the vicinity of the resonant nanoantenna was analyzed via lifetime distribution technique. The developed approach allows extraction of the velocity field with high accuracy in a broad range of possible flow velocities covering ranges significant for a variety of microfluidic applications. Additional design of antenna parameters can further increase the contribution to the flow-dependent component of the lifetime by providing stronger field localization and by the improvement of collection efficiency of spontaneously emitted photons with flow-modified lifetimes. The new approach of quantum sensing of motion allows mapping of dynamical processes beyond the diffraction limit via mapping of time-dependent near field interactions to the far field.

**Acknowledgments**